\documentclass[aps,prl,twocolumn]{revtex4-1}
\usepackage[dvips]{graphics,graphicx}
\usepackage{color}

\begin{document}
\title{Nucleation in finite topological systems during continuous \\  metastable quantum phase transitions}
\author{Oleksandr Fialko$^1$, Marie-Coralie Delattre$^{1}$, Joachim Brand$^1$ and Andrey R. Kolovsky$^{2,3}$}
\affiliation{$^1$Centre for Theoretical Chemistry and Physics and New Zealand Institute for Advanced Study, Massey University, Private Bag 102904 NSMC, Auckland 0745, New Zealand}
\affiliation{$^2$Kirensky Institute of Physics, 660036 Krasnoyarsk, Russia}
\affiliation{$^3$Siberian Federal University, 660041 Krasnoyarsk, Russia}

\date{\today}

\begin{abstract}
Finite topological quantum systems can undergo continuous metastable quantum phase transitions to change their topological nature. Here we show how to nucleate the transition between ring currents and dark soliton states in a toroidally trapped Bose-Einstein condensate. An adiabatic passage to wind and unwind its phase is achieved by explicit global breaking of the rotational symmetry.  This could be realized with current experimental technology.
\end{abstract}


\maketitle

Phase transitions have long been considered equilibrium phenomena of infinite systems that can involve interesting non-equilibrium nucleation dynamics, 
e.g.\ the formation of  topological defects \cite{KZM}. In recent years, quantum phase transitions have also been identified in excited states 
of nuclei as well as ultracold quantum gases
\cite{CarrBook,Caprio08,konotop09,Kana08} by the nonanalytic change in spectral properties, or their finite-system precursors, as a system parameter 
is changed. Such phase transitions are not manifested in equilibrium but they have dynamical consequences \cite{Caprio08,konotop09}. So far, little is 
known about the nucleation process.

Nucleation of equilibrium quantum phase transitions often relies on external symmetry breaking, which can occur {\em locally}.
An example is the Ising model  \cite{sachdev11}, where a tiny external magnetic field is sufficient for breaking the rotational symmetry of the 
spins and allowing the system to equilibrate. Similar mechanism were discussed in the context of vortex nucleation in Bose gases  \cite{dalib09,cohen08,david11}.
Here we theoretically investigate the quantum phase transition between metastable states of a Bose gas in a rotating toroidal trap identified in Ref.~\cite{Kana08}. 
The transition occurs between topological {\em vortex states} and non-topological {\em soliton states}. Vortex states of $N$ atoms correspond to ring currents and carry integer angular momentum per particle $L/(N\hbar)=J$. As the rotation frequency of the trap $\Omega$ is changed, the ground state  jumps between vortex states \cite{brand09} and thus the integer $J$ is a topological charge. The soliton states carry non-integer $L/(N\hbar)$. In mean-field theory they are approximated by dark solitons \cite{GP}, which carry a localized density notch and thus break
rotational symmetry in addition to changing the topological nature of the system, although no symmetry breaking is required for the phase transition 
in the quantum description of the finite system.

\begin{figure}[t]
\center
\includegraphics[width= 1\columnwidth, clip]{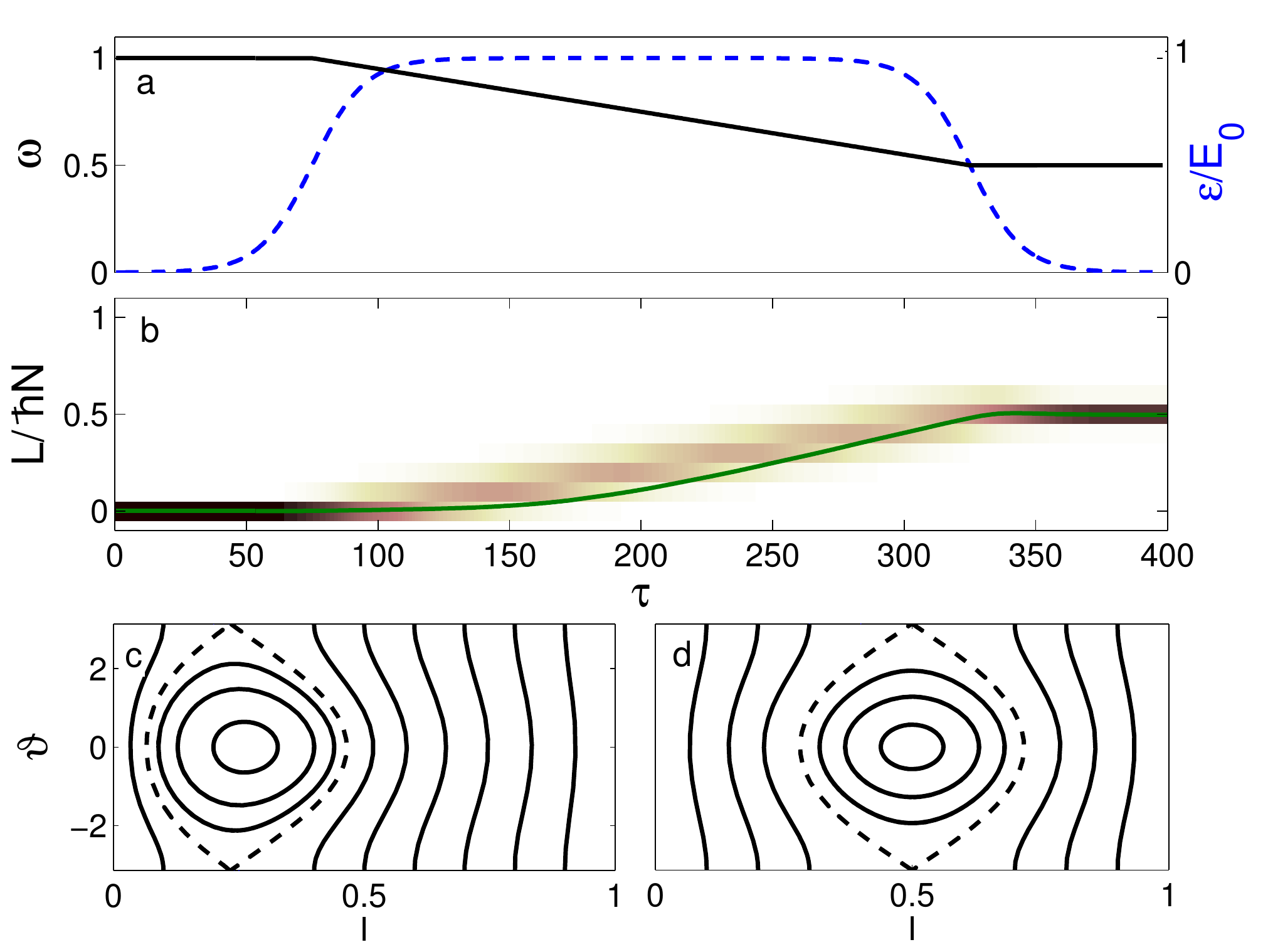}
\caption{Adiabatic passage from the ground to the dark soliton state.
(a) Protocol for variations of the perturbation parameter $\epsilon$ (dashed line) and 
the precession frequency $\omega$ (full line) for $\gamma=0.8 E_0$. 
(b) Angular momentum per particle
in the mean-field approximation (\ref{Eq.modelR}) over time (full line) and 
probability density of the quantum two-mode simulation with Hamiltonian (\ref{3}) on the same axes. White and black correspond to zero and maximum probability density, respectively.
Phase portraits of Eq.~(\ref{8}) 
are presented 
for $\omega=0.7$ (c) and $\omega=0.5$ (d).}
\label{fig1}
\end{figure}

\begin{figure}[t]
\center
\includegraphics[width= 1\columnwidth, clip]{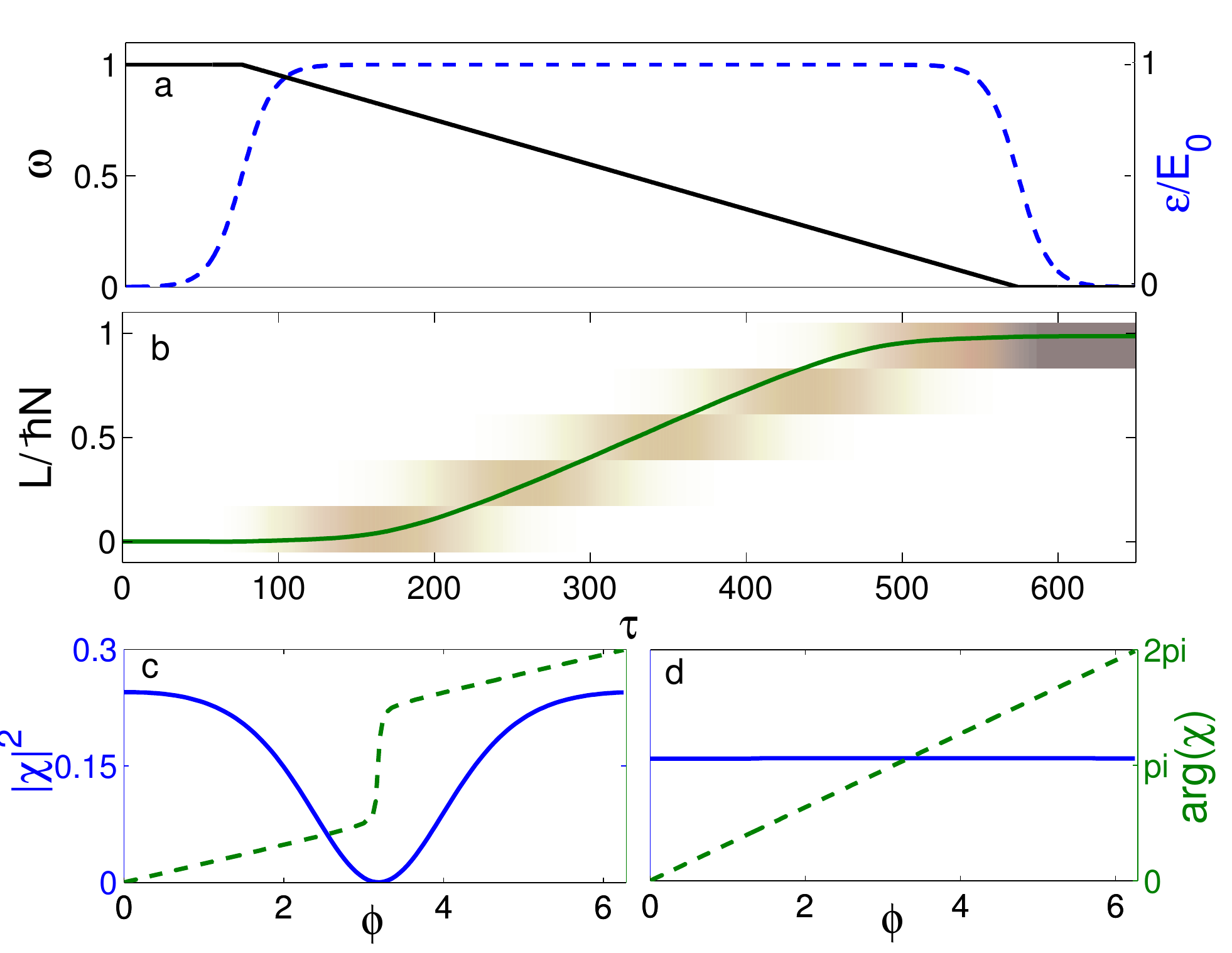}
\caption{Adiabatic passage from the ground state to the $J=1$ vortex state.
(a) Protocol for variations of the perturbation parameter $\epsilon$ (dashed line) and
the precession frequency $\omega$ (full line) for $\gamma=0.8 E_0$. 
(b) Angular momentum per particle
in the mean-field approximation (full line) and probability density of the corresponding quantum simulation  as in Fig.\ \ref{fig1}.
The dark soliton state (c) 
and the vortex state (d) represent the final states of
the corresponding mean field simulations.
Full lines represent density, dashed lines represent phases.}
\label{fig2}
\end{figure}

In this Letter, we show how to nucleate the phase transition by means of a {\em global} symmetry breaking 
potential, which creates an adiabatic passage through metastable states.
Although symmetry can be restored for a finite system, the passage naturally leads to the emergence of symmetry broken soliton solutions for large particle number $N$.
In order to maintain metastability, we  further find that local symmetry breaking needs to be avoided as it would lead to the unwanted thermalization of the excited states \cite{thermalization}. This is in stark contrast to equilibrium phase transitions, where thermalization is desirable.
The explicit global symmetry breaking is achieved by tilting the trap axis by some angle $\theta$ and rotating it with the frequency $\Omega$.
The frequency $\Omega$ is decreased from a maximum value $\Omega_\mathrm{initial}$ to a minimum value $\Omega_\mathrm{final}$ within a finite time interval, after which the tilt is decreased.
Under certain conditions angular momentum is established in the Bose gas up to  a dark soliton state or a vortex state
as seen in Figs.~\ref{fig1} and \ref{fig2}, respectively.
The protocol is counterintuitive as we decrease the rotation frequency of the system to increase its angular momentum.
It is reminiscent of the stimulated Raman adiabatic passage used in quantum optics to populate a metastable state  
through an intermediate level by a counterintuitive pulse sequence \cite{STIRAP}.

Interacting  bosonic atoms of mass $m$ confined to a rotating toroidal trap are described by the Hamiltonian 
\begin{equation}
\label{1} 
\widehat{H}=\int dx \hat{\Psi}^\dagger \left[-\frac{\hbar^2}{2m}\frac{\partial^2}{\partial x^2}
+\frac{U}{2}\hat{\Psi}^\dagger\hat{\Psi}+\epsilon\cos\left(\frac{x}{R}-\Omega t\right)\right]\hat{\Psi} \;,
\end{equation}
where $\hat{\Psi} \equiv \hat{\Psi}(x)$ is the atomic field operator, $U$ the bare interaction constant assumed repulsive ($U>0$),
and $\epsilon=mgR\sin(\theta)$ the strength of the symmetry breaking potential (where $mg$ is the gravitational force). 
The toroidal trap is approximated by a one-dimensional ring of the radius $R$. 

\begin{figure}[t]
\center
\includegraphics[width=1\columnwidth, clip]{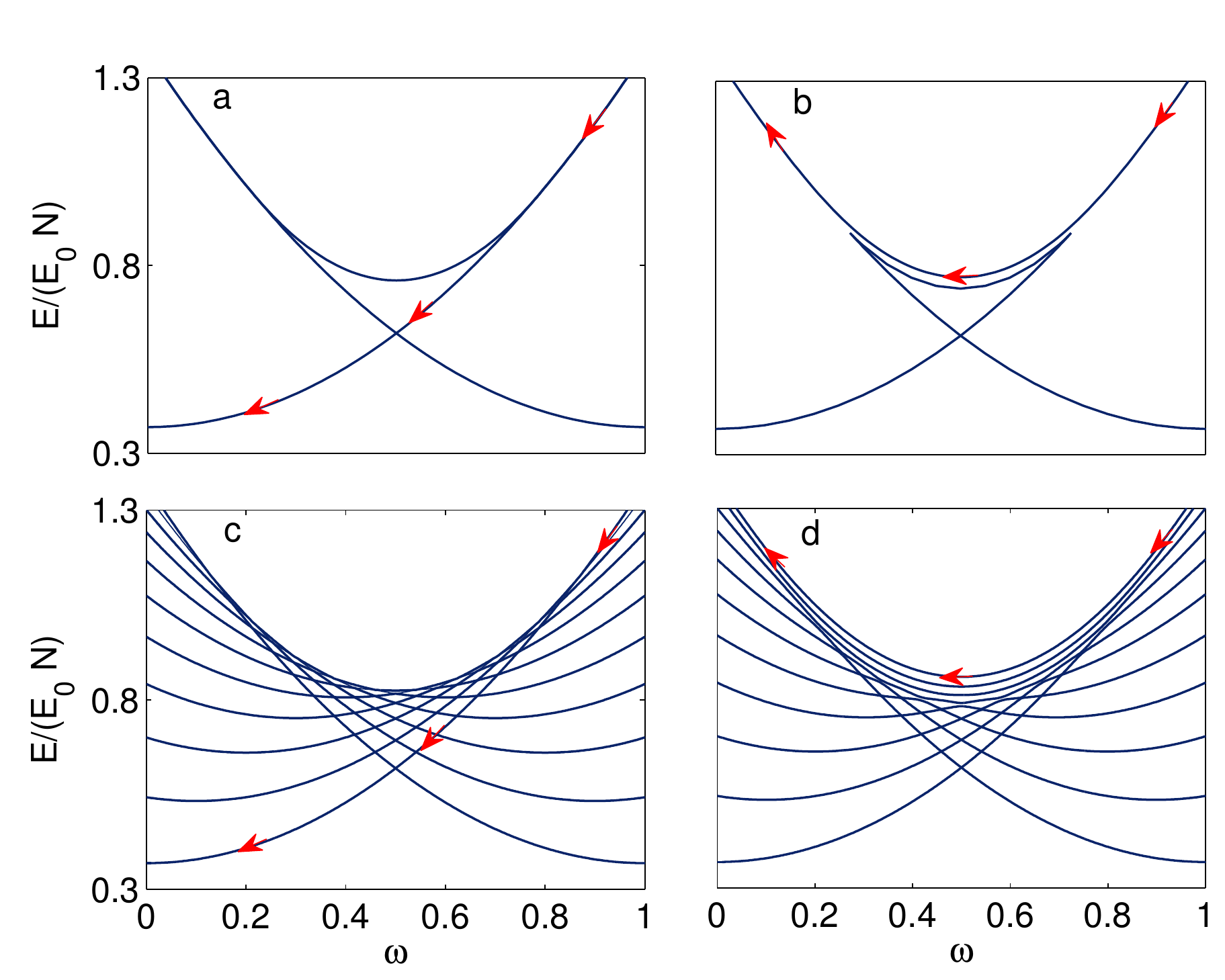}
\caption{
The energy of the GP equation (upper row) and 
the energy spectrum of the Hamiltonian (\ref{3}) in the 2-modes approximation (lower row). 
Parameters are $\gamma =0.8 E_0$, and $\epsilon=0$ (left column) and $\epsilon=0.04 E_0$ (right column). 
The arrows show the passage under adiabatic change of the frequency $\omega$.
Three distinct solutions exist in panel (a) between the bifurcation points $\omega_-\approx 0.13$ and $\omega_+\approx 0.87$.} 
\label{fig3}
\end{figure}

Let us first discuss the outlined problem in the mean-field approximation. Using a rotating coordinate frame with frequency $\Omega$ and 
introducing the scaled variables $\phi=x/R$, $\tau=tE_0/\hbar$, $\omega=\hbar\Omega/(2E_0)$, the Gross-Pitaevskii (GP) equation reads~\cite{GP}
\begin{equation}
i\frac{\partial\chi}{\partial \tau}=\left[-\left(\partial_\phi-i\omega\right)^2+2\pi\frac{\gamma}{E_0}|\chi|^2+\frac{\epsilon}{E_0} 
\cos(\phi)\right]\chi \;,
\label{Eq.modelR}
\end{equation}
where $\gamma=U(N-1)/(2\pi R)$ and $E_0=\hbar^2/(2mR^2)$ gives the relevant energy scale. The classical field $\chi$ is normalized as 
$\int_0^{2\pi}|\chi|^2 d\phi=1$ and it is periodic on $\phi$, $\chi(0)=\chi(2\pi)$. We apply Broyden's method \cite{broyden} 
to solve Eq.~(\ref{Eq.modelR}) numerically
for the stationary solutions $\chi(\tau,\phi)=\chi(\phi)e^{-i\mu\tau}$, where $\mu E_0$ is the chemical potential. 
For $\epsilon=0$ we present in Fig.~\ref{fig3}(a) three solutions corresponding to two vortex states $\chi(\phi)=e^{i J\phi}/\sqrt{2\pi}$ 
with $J=0$ and $J=1$, which are connected by a soliton branch. Finite $\epsilon$ opens a gap in the soliton branch, 
see Fig.~\ref{fig3}(b). 
The lower branch corresponds to a soliton sitting on the crest, while the upper branch connects to a soliton located in the trough of the external 
potential. In the following we will show that the lower branch is dynamically unstable, while the upper branch is stable. 
Thus an adiabatic passage through the upper branch is possible. Numerical simulations of the system dynamics on the basis of Eq.~(\ref{Eq.modelR}) 
confirm this expectation. The solid lines 
in Figs.~\ref{fig1}(b) and \ref{fig2}(b) show the momentum per atom as a function of 
time for the adiabatic protocols shown in the top panels. 
The final soliton and vortex states are shown in Figs.\ \ref{fig2}(c) and (d), respectively.

Mean-field theory thus presents the following picture: The  bifurcation points between the soliton and vortex branches seen in Fig.~\ref{fig3}(a) mark the transition between rotationally symmetric, and symmetry broken [Fig.\ \ref{fig2}(c)] phases. With finite $\epsilon$, the symmetry is formally broken everywhere and the bifurcation makes way to a swallowtail structure \cite{Mueller} in Fig.~\ref{fig3}(b). 
In contrast to the dynamics of continuous phase transitions in infinite systems where adiabaticity is always violated \cite{KZM}, an adiabatic passage is established (arrows in Fig.~\ref{fig3})  to change the symmetry properties of the initial state. Due to the global nature of the symmetry broken state with a single soliton in the whole ring there is no formation of domain structures with locally broken symmetry in further contrast to Refs.~\cite{KZM}.

We proceed with discussing and quantifying the necessary conditions for an adiabatic passage. 
Useful insight into the physics of the considered process is obtained within a two-mode approximation,  which is justified for small $\gamma\ll E_0$.  
Neglecting
all coefficients in the Fourier expansion except $k=0, 1$ of the function $\chi$, $\chi(\tau,\phi)=\sum_k b_k(\tau)\exp(ik\phi)$, 
the system dynamics is described by the effective Hamiltonian
\begin{equation}
\label{8} 
H_{\rm cl}= E_0(1-2\omega)I +\gamma I(1-I)+\epsilon\sqrt{I(1-I)}\cos\vartheta \;,
\end{equation}
where $I=|b_1|^2$ and $\vartheta$ is the relative phase for the amplitudes $b_0$ and $b_1$. The phase portrait of (\ref{8}) contains a stability island 
around the elliptic point $(I,\vartheta)=(I^*,0)$ 
(cf. Fig.~\ref{fig1}c, d), where $I^*\approx 1/2 + (1/2+\omega)E_0/\gamma$ depends linearly on $\omega$ and reaches values of zero and unity for 
\begin{equation}\label{omegapm}
\omega_{\pm} = \frac{1}{2}\pm\frac{\gamma}{2E_0}, 
\end{equation}
 respectively.
We note that in the lab frame this stability island corresponds to a nonlinear resonance. Thus the adiabatic passage has a simple physical interpretation: 
by ramping $\epsilon$ to a finite value we capture the system into the nonlinear resonance, transport it to any desired value of $I$ by adiabatically 
changing the frequency $\omega$ from $\omega_\mathrm{initial} \gtrsim \omega_+$ to $\omega_\mathrm{final}$, and release the system by ramping $\epsilon$ back to zero. The necessary condition 
for the time scale of this process is $\Delta t \gg \Omega_s^{-1}$, where $\Omega_s\approx \sqrt{\gamma\epsilon}/\hbar$ is the frequency of small oscillations
near the elliptic point of the stability island  
(the approximate sign is replaced by an equal sign for $\omega=1/2$).

For stronger nonlinearity, $\gamma>E_0$, when the two-mode approximation is not justified, the adiabaticity conditions and stability
can be studied using the Bogoliubov approach, which linearizes the time dependent GP equation (\ref{Eq.modelR}) \cite{GP}.
This leads to the eigenvalue problem
\begin{eqnarray}
\nonumber
\lambda_i u_i=(\hat{H}_{GP}-\mu+2\pi \gamma/E_0 |\chi|^2)u_i+2\pi \gamma/E_0 \chi^2 v_i, \\
-\lambda_i v_i=(\hat{H}_{GP}-\mu+2\pi \gamma/E_0 |\chi|^2)v_i+2\pi \gamma/E_0 \chi^{\ast 2} u_i.
\end{eqnarray}
Here, $\chi=\chi_\omega(\phi)$ is the stationary solution for the upper soliton branch in  Fig.~\ref{fig3}(b) and $\hat{H}_{GP}$ 
is given in the square brackets in Eq.~(\ref{Eq.modelR}). For the lower branch we find a single imaginary eigenvalue $\lambda$,
which indicates that the solutions on this branch are dynamically unstable \cite{GP}.
The Bogoliubov analysis for the upper branch indicates that it is dynamically stable and that there is a gap to the lowest lying elementary excitation. 
The corresponding frequency $\Omega_s = \lambda E_0/\hbar$ has the asymptotic behavior
\begin{equation}
\Omega_s=\left\{
\begin{array}{ccc}
\sqrt{\gamma\epsilon}/\hbar &,& \gamma \ll E_0 \\
\sqrt{E_0\epsilon}/\hbar &,& \gamma \gg E_0 
\end{array} 
\right. \;.
\label{Eq.OmB}
\end{equation}
For small nonlinearities $\gamma\ll E_0$ this result agrees with the previous two-mode analysis.

The large $\gamma$ result can be understood by noting that the soliton is a localized object. 
In this case $\Omega_s$ is interpreted as the frequency of small oscillations of the soliton around a stationary point.
Indeed, it can be shown that the small oscillations of the soliton around the stationary 
points are described by the equation \cite{konotop} 
\begin{equation}
m\left(\frac{\partial\phi_s}{\partial t}\right)^2\pm\frac{\epsilon}{2R^2}\phi_s^2\approx {\rm const} \;,
\end{equation}
which is valid for $\Omega\approx E_0/\hbar$. The $\pm$ sign corresponds to a soliton sitting at the trough or the crest of the potential, 
respectively. While the minus sign describes an unstable situation (lower soliton branch in Fig.~\ref{fig3}b), the plus sign
describes  harmonic oscillations of a particle with mass $2m$. 
The soliton located in the trough is thus dynamically stable, with the frequency of small oscillations $\Omega_s=\sqrt{\epsilon/(2mR^2)}= \sqrt{E_0\epsilon}/\hbar$.

In practice, the adiabaticity condition $\Delta t\gg \Omega_s^{-1}$ discussed above should be made even stronger by requiring that accelerated rotation should displace the soliton by less than the {the soliton core size, which is of the order of} the healing length $\xi=\sqrt{\hbar^2/2m\gamma}$. This yields the condition
\begin{equation}
\frac{\partial\Omega}{\partial t} \ll \frac{\xi}{R}\frac{E_0 \epsilon}{\hbar^2}.
\label{Eq.adiabMF}
\end{equation}  
The numerical simulation of the system dynamics confirms the estimate (\ref{Eq.adiabMF}). 
For the parameters of  Fig.~\ref{fig1} the right hand side of (\ref{Eq.adiabMF}) is $\approx 0.04(E_0/\hbar)^2$. 
It is seen in Fig.~\ref{fig1} and \ref{fig2} that for a ten times smaller rate $\dot{\Omega}=0.004(E_0/\hbar)^2$ the dark soliton and the vortex 
states (green solid lines) are reached. 

The mean-field analysis presented so far can be substantiated by a quantum analysis. Using scaled variables, a rotating coordinate frame, 
and expanding the field operators in the angular-momentum basis, the Hamiltonian of the system takes the form of the Lieb-Liniger
Hamiltonian \cite{lieb-liniger} with an external symmetry-breaking potential under periodic boundary conditions 
\begin{eqnarray}
\nonumber
\widehat{H}=E_0\sum_k (k-\omega)^2
\hat{b}_k^\dagger\hat{b}_k +
\frac{\epsilon}{2}\sum_k\left(
\hat{b}^\dagger_{k+1}\hat{b}_{k}+h.c. \right)\\
+\frac{U}{4\pi R}\sum_{k_1,k_2,k_3,k_4}
\hat{b}_{k_1}^\dagger\hat{b}^\dagger_{k_2}\hat{b}_{k_3}\hat{b}_{-k_3-k_2-k_1},
\label{3} 
\end{eqnarray}
We are interested in the time evolution with initial condition given by the nonrotating ground state (vortex state with $J=0$). 
The soliton states have definite angular momentum for $\epsilon =0$ and, within the validity of the two-mode approximation, are given by 
\begin{equation}
\label{5} 
|\psi_l\rangle=(\hat{b}_0^{\dagger})^{N-l}(\hat{b}_1^{\dagger})^{l}|{\rm vac}\rangle . 
\end{equation}
Truncating the Fock basis to the states 
 (\ref{5}), the Hamiltonian (\ref{3}) is a tri-diagonal $(N+1)\times (N+1)$ matrix. Fig.~\ref{fig3}(c) shows the spectrum 
of this matrix for $N=10$,  $\gamma=0.8 E_0$, $\epsilon=0$, and $0\le\omega\le1$. Note that for $U=0$ all levels would cross at one point at $\omega=1/2$. 
Finite interactions remove this degeneracy, leading to the appearance of a caustic in the level crossing pattern. The spectrum of the system 
for a finite $\epsilon=0.04E_0$ is shown in Fig.~\ref{fig3}(d) \cite{Korsch}. Now all level crossings are substituted by avoided crossings. 
The Hamiltonian in the two mode model can be recast to take the form of a quantized version of the
the effective Hamiltonian (\ref{8}), where the action variable $I$ is now associated with the operator $\hat{I}=-i(1/N)\partial/\partial\vartheta$ and $1/N$ plays the role of the Planck constant. 
In particular, we find that the transition frequency between the caustic levels is given by  $\Omega_s$ of Eq.~(\ref{Eq.OmB}). 

Within the quantum model we can find a condition for the frequency  $\omega_\mathrm{initial}$ from the crossing point of the two energy levels that are lowest for $\omega=0$ in Fig.~\ref{fig3}(c). In the two mode approximation, we obtain $\omega_+$ of Eq.~(\ref{omegapm}), the same result as from the mean-field analysis.
A more thorough comparison of the upper and lower rows in Fig.~\ref{fig3} indicates that energies of the 
soliton states calculated within the mean-field approximation are shifted in the negative direction compared 
to the quantum calculation. This is a manifestation of the occupation of additional angular momentum modes outside the two-mode model. 
In order to extend the result for $\omega_+$ beyond the two-mode approximation, we consider the rotation frequency 
$\omega_{l}=\sqrt{l^2 + 2\gamma/ E_0}/2$ where the Bogoliubov phonon with angular momentum $l\hbar>0$ acquires zero energy. 
The expression for $\omega_+ \equiv \omega_{1}$ thereby generalizes Eq.\ (\ref{omegapm}) to larger values of $\gamma$. Generally, it looks 
possible to excite the vortex state with topological charge $J$ with the condition $\omega_{J}<\omega_\mathrm{initial}<\omega_{J+1}$ 
\cite{simulat} through metastable states with $J$ dark solitons \cite{Kana08}.

It is interesting to simulate the dynamics of the quantum system (\ref{3}) for the same protocols as were used in the mean-field simulations. 
Due to exponential proliferation  of the Hilbert space with $N$, this can be done only for a small number of atoms $N\sim 10$. The density plots of
Figs.~\ref{fig1}(b) and \ref{fig2}(b) show the dynamics of the system (\ref{3}) in the two-mode approximation for $N=10$ and $\gamma=0.8 E_0$. 
(We have checked that for these value of the interaction constant and number of particles the result remains unchanged if we use a four-mode 
approximation.) The gray scale encodes populations of the Fock states (\ref{5}). It is seen in Fig.~\ref{fig1} and \ref{fig2} 
that our protocols almost entirely 
populate the target excited state  $|\psi_{N/2}\rangle$ (Fig.~\ref{fig1}(b)) and $|\psi_{N}\rangle$ (Fig.~\ref{fig2}(b)), 
respectively, with the probabilities $0.93$ and $1$. 

The quantum model (\ref{3}) reveals why global symmetry breaking is effective for nucleating the metastable 
phase transition: The long wavelength perturbing potential proportional to $\epsilon$ couples neighboring soliton states 
(\ref{5}) effectively and thus creates the avoided crossings that provide the adiabatic passage. 
A short wavelength, {\em local} symmetry breaking perturbation, in contrast, would tentatively couple to excited states with larger angular momentum difference, generate drag, and thus compromise the adiabatic passage. For a large system ($R\gg \xi$) this is expected from both quantum \cite{brand09} and mean-field \cite{Hakim} consideration of a localised impurity moving at supersonic velocity $\Omega R > \Omega_1 R = c +{\cal O}(R^{-1})$, where $c=\sqrt{\gamma/m}$ is the speed of sound.

There is still a crucial difference between the 
quantum and mean field models: While the soliton states of the GP approximation break the rotational symmetry, the corresponding 
quantum states  (\ref{5}) do not. Instead, they correspond to fragmented condensates, where both the occupation of the $k=0$ and the $k=1$ mode can become large.
Whether we obtain a fragmented condensate with preserved rotational 
symmetry or a condensate with broken symmetry, depends on the rate for changing the parameter $\epsilon$ at the very end of 
the passage. Simple calculation reveals that adiabatic condition is fulfilled if the rate satisfies $d\epsilon/dt\ll\gamma^2/(2\hbar N)$ \cite{schiff}. 
For large $N$ it becomes more difficult to restore symmetry, since the time scales required for restoring the symmetry become large.
When $\epsilon$ is changed back to zero in finite time, the system `sits' 
on the caustic where the density of states tends to infinity if $N\rightarrow\infty$.  
That a large particle number brings a qualitative change is familiar from the general principle of symmetry breaking in condensed matter \cite{anderson}. In contrast to conventional phase transitions, however, where symmetry breaking occurs spontaneously, the metastable quantum phase transition discussed in this Letter requires an explicit breaking of its symmetry.

So far we have mostly considered the case of moderate nonlinearity, where the two-mode approximation is justified. 
In typical experiments with BECs, larger nonlinearities beyond the validity of the two-mode approximation are relevant. We have 
verified within the mean-field theory that adiabatic passage for stronger nonlinearities is possible as well \cite{simulat}. 
We estimate the time scale for the adiabatic passage to a dark soliton state 
using Eq.~(\ref{Eq.adiabMF}) and setting $\epsilon\sim \gamma$ as $\Delta t \gg m R^2/\hbar$.
In the recent experiment with $^{23}$Na \cite{hill11}, $R\sim 20\mu m$. This gives $\Delta t \gg  1 s$.  For lighter atoms, e.g. $^{7}$Li, and smaller radius, e.g. $R\sim 10\mu m$, one can achieve the condition $\Delta t \gg 0.1 s$ and $\theta\approx 0.5^{\rm o}$. Therefore,  time scales of the order of $1s$ will provide an adiabatic passage. 

In conclusion, we have proposed an experimental scheme to nucleate a metastable quantum phase transition in an ultra-cold Bose gas by an adiabatic passage.
We have shown that this is achieved by an explicit breaking of the global rotational symmetry. The procedure generalizes nucleation procedures, which change the ground state symmetries in the course of ordinary continuous phase transitions. 

We acknowledge stimulating discussions with Jean-Sebasti\'en Caux  and Lincoln Carr. AK thanks Massey University for hospitality.
OF, MCD and JB were supported by the Marsden Fund (contract No.\ MAU0910) administrated by the Royal Society of New Zealand. 


\end{document}